\documentclass[12pt]{article}

\catcode`\@=11
\@addtoreset{equation}{section}

\global\arraycolsep=2pt 
\usepackage{mathrsfs, amsbsy, amssymb, latexsym, amsfonts, amsmath, cite} 
\usepackage{graphicx, color, amscd}

\newcommand{\alg}[1]{\mathfrak{#1}}

\newcommand{\wt}[1]{\widetilde{#1}}

\newcommand{\al}{\alpha}
\newcommand{\be}{\beta}

\newcommand{\de}{\delta}

\newcommand{\la}{\lambda}

\newcommand{\no}{\nonumber}
 
\newcommand{\ran}{\rangle}
\newcommand{\lan}{\langle}
 
\allowdisplaybreaks

\begin{document}

\begin{flushright}
\parbox{4cm}
{KUNS-2477 \\ 
ITP-UU-14/05 \\ 
SPIN-14/05}
\end{flushright}

\vspace*{1.5cm}

\begin{center}
{\Large\bf Jordanian deformations of the AdS$_5\times$S$^5$ superstring}
\vspace*{2.5cm}\\
{\large Io Kawaguchi$^{\ast}$\footnote{E-mail:~io@gauge.scphys.kyoto-u.ac.jp}, 
Takuya Matsumoto$^{\dagger}$\footnote{E-mail:~t.matsumoto@uu.nl} 
and Kentaroh Yoshida$^{\ast}$\footnote{E-mail:~kyoshida@gauge.scphys.kyoto-u.ac.jp}} 
\end{center}
\vspace*{0.25cm}
\begin{center}
$^{\ast}${\it Department of Physics, Kyoto University \\ 
Kyoto 606-8502, Japan.} 
\vspace*{0.25cm}\\ 
$^{\dagger}${\it Institute for Theoretical Physics and Spinoza Institute, 
Utrecht University, \\ 
Leuvenlaan 4, 3854 CE Utrecht, The Netherlands.} 
\end{center}
\vspace{1cm}

\begin{abstract}
We consider Jordanian deformations of the AdS$_5\times$S$^5$ superstring action. 
The deformations correspond to non-standard $q$-deformation. 
In particular, it is possible to perform partial deformations, for example, only for the S$^5$ part. 
Then the classical action and the Lax pair are constructed with a linear, twisted and extended 
$R$ operator. It is shown that the action preserves the $\kappa$-symmetry. 
\end{abstract}

\setcounter{footnote}{0}
\setcounter{page}{0}
\thispagestyle{empty}

\newpage

\section{Introduction}

One of the fascinating topics in string theory is the AdS/CFT correspondence 
\cite{M,GKP,W}. The most well-studied example is the duality between type IIB superstring on 
the AdS$_5\times$S$^5$ background \cite{MT} 
(often called the AdS$_5\times$S$^5$ superstring) and the $\mathcal{N}=4$ $SU(N)$ 
super Yang-Mills (SYM) theory in four dimensions (in the large $N$ limit). 
It has been revealed that the integrable structure exists behind the duality 
and it plays a fundamental role in testing the correspondence of physical quantities  
(For a comprehensive review, see \cite{review}). 

\medskip 

Our interest here is the integrability in the string-theory side. The classical integrable structure 
of the AdS$_5\times$S$^5$ superstring is closely related to 
the $Z_4$-grading property of the supercoset \cite{BPR}\footnote{For the classical integrability based on 
the Roiban-Siegel formulation \cite{RS}, see \cite{Hatsuda}.}, 
\[
PSU(2,2|4)/[SO(1,4)\times SO(5)]\,.
\]
The supercosets with the grading property are classified, 
including the stringy conditions \cite{Zarembo-symmetric}. 

\medskip 

The next is to consider integrable deformations. 
There are two approaches, the one is based on 1) deformed S-matrices 
and the other is based on  2) deformed target spaces. 
For the first approach, the deformed S-matrices are constructed in a mathematically 
well-defined way \cite{BK,BGM,HHM,dLRT,Arutyunov}, but the corresponding geometry of target space is unclear. 
In the second direction, the classical integrable structure has been well studied for 
three-dimensional examples such as squashed S$^3$ 
(For the classic works and the recent progress, 
see \cite{Cherednik,FR,BFP} and \cite{KY,KYhybrid,ORU,KMY-QAA,
KMY-monodromy,KOY}, respectively)  
and warped AdS$_3$ \cite{KY-Sch, Jordanian-KMY,Kame}. The deformed geometries 
are represented by non-symmetric cosets \cite{SYY} and there is no general prescription 
to argue the integrability. 
For generalizations to higher dimensions, see \cite{BR,DMV}.  
In particular, the method utilized in \cite{DMV} is based on Yang-Baxter sigma models \cite{Klimcik}. 
The standard $q$-deformation of $\alg{su}(2)$ \cite{Drinfeld1,Drinfeld2,Jimbo} 
and the affine extension are also presented \cite{KYhybrid,DMV} and \cite{KMY-QAA}, respectively.  

\medskip 

Recently, a $q$-deformed AdS$_5\times$S$^5$ superstring action was constructed \cite{DMV2} 
by generalizing the result in \cite{DMV}. 
Then the bosonic part of the action was determined and, by using the action, 
the world-sheet S-matrix of bosonic excitations was computed in \cite{ABF}. 
The resulting S-matrix exactly agrees with the $q$-deformed S-matrix in the large 
tension limit. Thus the two approaches will now be related each other 
and there are many directions to study $q$-deformations of the AdS$_5\times$S$^5$ superstring.  

\medskip 

In this paper, we consider how to twist the $q$-deformed AdS$_5\times$S$^5$ superstring action. 
This twisting is regarded as a non-standard $q$-deformation. Indeed,  
it would also be seen as a higher-dimensional generalization of 3D Schr\"odinger sigma models 
in which $q$-deformed Poincare algebra \cite{q-Poincare,Ohn} and its infinite-dimensional extension 
are realized as shown in a series of works \cite{KY-Sch,Jordanian-KMY}. 
In particular, it is possible to perform partial deformations, for example, only for the S$^5$ part. 
It would make the resulting geometry much simpler. Some extensions of the twisted $R$ operators 
are also discussed. 
Then the classical action and the Lax pair are 
constructed with a linear, twisted and extended $R$ operator. 
It is shown that the action preserves the $\kappa$-symmetry.   

\medskip 

The paper is organized as follows. Section 2 is a short review of the $q$-deformed 
AdS$_5\times$S$^5$ action. Section 3 describes how to twist the $q$-deformed action.  
Then we construct the Jordanian deformed action of the AdS$_5\times$S$^5$ superstring 
preserving the $\kappa$-symmetry. The Lax pair is also presented. 
Section 4 is devoted to conclusion and discussion. Appendix A describes the notation of 
the superconformal generators. In Appendix B, the notation of the classical $R$-matrix is explained. 
A general prescription to twist the classical $r$-matrix for the standard $q$-deformation 
of Drinfeld-Jimbo type is also provided.

\section{A review of the $q$-deformed AdS$_5\times$S$^5$ superstring}

In this section, we will give a short review of the $q$-deformed AdS$_5\times$S$^5$ 
superstring action constructed in \cite{DMV2}, using the notation therein. 

\subsection{The linear $R$ operator}

A key ingredient in the construction is the classical $R$-matrix, 
which is a linear map $R:\alg{g}\to \alg{g}$ over a Lie algebra $\alg{g}$ satisfying 
the modified classical Yang-Baxter equation (mCYBE);   
\begin{eqnarray}
\left[R(M),R(N)\right]-R\left(\left[R(M),N\right]+\left[M,R(N)\right]\right)=-c^2\left[M,N\right]\,, 
\label{mCYB}
\end{eqnarray}
where $M,N \in \alg{g}$ and $c$ is a complex parameter. 
When $c\neq 0$, the parameter is regarded as a scaling of the $R$-matrix 
and could be normalized as $c=1$. 
When $c=0$, the mCYBE is nothing but the classical Yang-Baxter equation (CYBE). 

\medskip 

The standard $q$-deformation of the superstring action presented in \cite{DMV2} is 
described by the following $R$-matrix, 
\begin{align}
R(E_{ij})=
\begin{cases} + c E_{ij} \quad \text{for}\quad i<j \\ 
-c E_{ij} \quad \text{for} \quad i>j 
\end{cases}  
\qquad \text{and} \qquad R(E_{ii})=0\,, 
\label{triR}
\end{align}
where $E_{ij}$ ($i,j = 1, \cdots, 8$) are the $\alg{gl}(4|4)$ generators. 
For the standard notation of the superconformal generators, see Appendix \ref{app:sconf}. 
The parity of the indices is given by 
$\bar i = 0$ for $i=1,\cdots, 4$ and $\bar i = 1$ for $i=5,\cdots, 8$.  
The associated tensorial $r$-matrix is  
\begin{align}
r_{\rm DJ} = c \sum_{1\leq i< j \leq 8} E_{ij} \wedge E_{ji}\,(-1)^{{\bar i}\,{\bar j}}\,,  
\label{triR12}
\end{align}
where the super skew-symmetric symbol is introduced as  
\begin{align}
E_{ij}\wedge E_{kl} \equiv E_{ij}\otimes E_{kl} 
- E_{kl}\otimes E_{ij} (-1)^{(\bar i +\bar j)(\bar k +\bar l)}\,.  
\end{align}
The relations between the linear $R$ operator and 
the tensorial notation $r$ are summarized in Appendix \ref{app: classialR}. 
The classical $r$-matrix given in (\ref{triR12}) describes the standard 
$q$-deformation of Drinfeld-Jimbo (DJ) type \cite{Drinfeld1,Drinfeld2,Jimbo}.

\subsection{The classical action and the Lax pair}

With the help of the linear $R$ operator defined in \eqref{triR}, 
the $q$-deformed classical action $S$ is given 
by\footnote{Here we have normalized the parameter as $c=1$ in \eqref{triR}. }  
\begin{eqnarray}
S = -\frac{(1+\eta^2)^2}{2(1-\eta^2)} \int^{\infty}_{-\infty}\!\!d\tau\int^{2\pi}_0\!\!d\sigma\,P_-^{\alpha\beta}\, 
{\rm Str}\left(A_{\alpha}\, d\circ \frac{1}{1-\eta R_g\circ d} (A_{\beta}) \right)\,. 
\label{q-action}
\end{eqnarray}
Here $\tau$ and $\sigma$ are time and spatial coordinates of the string world-sheet 
and the periodic boundary condition is imposed for the $\sigma$ direction. The real constant  
$\eta \in [0,1)$ measures the deformation\footnote{Since the deformation is measured by $\eta$\,, 
it is often called ``$\eta$-deformation''. On the other hand, $\eta$ 
is related to the $q$ parameter of the standard $q$-deformation by Drinfeld-Jimbo 
\cite{Drinfeld1,Drinfeld2,Jimbo} as shown in \cite{DMV}. Hence we will refer this deformation as to 
$q$-deformation, following \cite{DMV2}.}. The super Maurer-Cartan one-form 
$A_{\alpha}$ is defined as 
\[
A_{\alpha} \equiv g^{-1}\partial_{\alpha} g\,, \qquad g \in SU(2,2|4)\,,
\]
and $A_{\alpha}$ takes the value in the Lie superalgebra $\mathfrak{su}(2,2|4)$\,. 
The action of the $R$-matrix \eqref{triR} on $A_{\alpha}$ is induced from $\alg{gl}(4|4)$
by imposing a suitable reality condition. 
Note that $A_{\alpha}$ automatically satisfies the flatness condition, 
\begin{eqnarray}
\mathcal{Z} \equiv \frac{1}{2}\epsilon^{\alpha\beta}
\left(\partial_{\alpha}A_{\beta} - \partial_{\beta}A_{\alpha} + [A_{\alpha},A_{\beta}]\right) =0\,. \label{q-flat-A}
\end{eqnarray} 
The projection operators $P_{\pm}^{\alpha\beta}$ are defined as 
\[
P_{\pm}^{\alpha\beta} \equiv \frac{1}{2}\left(\gamma^{\alpha\beta} \pm \epsilon^{\alpha\beta}\right)\,.
\]
Then operators $d$ and $\tilde{d}$ are linear combinations of the projection operators $P_i~(i=1,2,3)$\,, 
\begin{eqnarray}
d \equiv P_1 + \frac{2}{1-\eta^2}\,P_2 -P_3\,, \qquad 
\tilde{d} \equiv -P_1 + \frac{2}{1-\eta^2}\,P_2 + P_3\,.
\end{eqnarray}
The symbol $R_g$ indicates a chain of the adjoint operation and the linear $R$ operation, 
\begin{eqnarray}
R_g(M) \equiv Ad_g^{-1}\circ R\circ Ad_g(M) = g^{-1}R(gMg^{-1})g\,. 
\end{eqnarray}
Note that the usual AdS$_5\times$S$^5$ superstring action is reproduced from (\ref{q-action}) 
when $\eta =0$\,. 
For a pedagogical review of the undeformed AdS$_5\times$S$^5$ superstring, see \cite{AF-review}\,.

\medskip 

It is convenient to introduce the following notations,
\begin{eqnarray}
J_{\alpha} \equiv \frac{1}{1-\eta R_g\circ d}\,(A_{\alpha})\,, \quad 
\widetilde{J}_{\alpha} \equiv \frac{1}{1+\eta R_g\circ \tilde{d}}\,(A_{\alpha})\,, 
\quad J_-^{\alpha} \equiv P_-^{\alpha\beta}J_{\beta}\,, \quad 
\widetilde{J}_+^{\alpha} \equiv P_+^{\alpha\beta}\widetilde{J}_{\beta}\,. 
\end{eqnarray}
Then the equations of motion are written in a simpler form, 
\begin{eqnarray}
\mathcal{E} = d(\partial_{\alpha} J_-^{\alpha}) + \tilde{d}(\partial_{\alpha}\widetilde{J}_+^{\alpha}) 
+ [\widetilde{J}_{+\alpha}, d (J_-^{\alpha})] + [J_{-\alpha},\tilde{d}(\widetilde{J}_+^{\alpha})] = 0\,. 
\label{q-eom}
\end{eqnarray}
The Lax pair is given by 
\begin{eqnarray}
L_+^{\alpha} &=& \widetilde{J}_+^{\alpha (0)} + \lambda \sqrt{1+\eta^2}\,\widetilde{J}_+^{\alpha (1)} 
+ \lambda^{-2} \left(\frac{1+\eta^2}{1-\eta^2}\right)\widetilde{J}_+^{\alpha (2)} 
+ \lambda^{-1}\sqrt{1+\eta^2}\,\widetilde{J}_+^{\alpha (3)}\,, \nonumber \\ 
M_-^{\alpha} &=& J_-^{\alpha (0)} + \lambda\sqrt{1+\eta^2}\,J_-^{\alpha (1)} 
+ \lambda^{2} \left(\frac{1+\eta^2}{1-\eta^2}\right)J_-^{\alpha (2)} 
+ \lambda^{-1}\sqrt{1+\eta^2}\,J_-^{\alpha (3)}\,, 
\label{q-Lax}
\end{eqnarray}
where $\lambda$ is the spectral parameter that takes a complex value. 
The flatness condition (\ref{q-flat-A}) can be rewritten in terms of $J_-^{\alpha}$ and $\widetilde{J}_+^{\alpha}$ like 
\begin{eqnarray}
\mathcal{Z} = \partial_{\alpha}\widetilde{J}_+^{\alpha} - \partial_{\alpha}J_-^{\alpha} 
+[J_{-\alpha},\widetilde{J}_+^{\alpha}] + \eta^2 [d(J_{-\alpha}), \tilde{d} (\widetilde{J}_+^{\alpha})] 
+ \eta R_g(\mathcal{E})=0\,. 
\label{q-flat}
\end{eqnarray}
With the definition $\mathcal{L}_{\alpha} \equiv L_{+\alpha} + M_{-\alpha}$\,, the zero-curvature 
condition
\begin{eqnarray}
\partial_{\alpha}\mathcal{L}_{\beta} - \partial_{\beta}\mathcal{L}_{\alpha} + [\mathcal{L}_{\alpha},\mathcal{L}_{\beta}] =0
\label{zero}
\end{eqnarray}
is equivalent to the equations of motion given in (\ref{q-eom}) and the flatness condition (\ref{q-flat})\,. 
For the $\kappa$-symmetry argument, see \cite{DMV2}.

\section{Jordanian deformations of the AdS$_5\times$S$^5$ superstring} 

In this section we shall consider Jordanian deformations of the AdS$_5\times$S$^5$ superstring action. 
The deformations correspond to a non-standard $q$-deformation and contain twists of the linear 
$R$ operator. The twist procedure is realized as an adjoint operation for the linear $R$ operator 
with an arbitrary bosonic root. Also, see Appendix \ref{app: classialR}.  

\medskip 

We first explain how to construct Jordanian $R$ operators
by twisting the linear $R$ operator used in the $q$-deformed 
AdS$_5\times$S$^5$ superstring action (\ref{q-action})\,. 
There are two remarkable features of Jordanian $R$ operators. 
The first is that they satisfy CYBE rather than mCYBE \eqref{mCYB}.  
The second is the nilpotency of them. That is 
\begin{eqnarray}
&&\left[R(M),R(N)\right]-R\left(\left[R(M),N\right]+\left[M,R(N)\right]\right)=0   
\label{CYB}\,, \\
&&R^n(M)=0 \qquad \text{for}\quad n\geq 3\,, 
\end{eqnarray}
for $M,N\in \alg{g}$\,.  

\medskip 

Then, by using the Jordanian $R$ operators, the Jordanian deformed 
action with the $\kappa$-symmetry and the Lax pair are presented.

\subsection{Jordanian $R$ operators from twists and their extension} 

We shall give a description to twist the linear $R$ operator 
for basic examples of Jordanian $R$ operators here. 
Then some extensions of twisted $R$ operators are discussed. 

\medskip 

First of all, note that the classical $r$-matrix of Drinfeld-Jimbo type \eqref{triR12} 
has the vanishing Cartan charges,
\begin{align}
[\Delta(E_{ii}), r_{\rm DJ}]=0\qquad\text{for} \qquad
i=1, \cdots, 8\,,
\end{align} 
where the coproduct is given by 
\[
\Delta(X)=X\otimes1+1\otimes X \qquad \mbox{for} ~~X \in \alg{g}\,.
\label{copro}
\]     

\medskip 

On the other hand, one may introduce a classical $r$-matrix which has non-zero Cartan charges
for the deformation of AdS$_5\times$ S$^5$ superstring. 
In this sense, we refer those as to Jordanian $r$-matrices.     
In general, such an $r$-matrix can be constructed by a twist of $r_{\rm DJ}$ 
with an arbitrary bosonic root $E_{ij}$ with $i<j$\,,  
\begin{eqnarray}
r_{\rm tw}^{(i,j)} \equiv [\Delta (E_{ij}),r_{\rm DJ}]\,. 
\end{eqnarray} 
One may also consider twists by negative bosonic roots $E_{ij}~(i>j)$, 
but the corresponding $r$-matrix has the same property 
because $\alg{gl}(4|4)$ algebra enjoys the automorphism 
\begin{align}
E_{ij}\mapsto E_{9-j, 9-i}\,.   
\end{align}
Thus positive roots $E_{ij}~(i<j)$ are enough for our later argument. 
The twisted, linear $R$ operator is defined as 
\begin{eqnarray}
R_{\rm tw}^{(i,j)} (X) &\equiv& \langle r_{\rm tw}^{(i,j)}, 1\otimes X \rangle \nonumber\\ 
&=& [E_{ij},R(X)] -R([E_{ij},X]) \qquad \mbox{for}~~X \in \mathfrak{g}\,.  
\label{Rr} 
\end{eqnarray}
It is straight forward to read off the $R$ operator
from the tensorial $r$-matrix via \eqref{Rr} and inner product \eqref{inp}. 

\medskip 

So far, we have constructed the Jordanian 
$R$ operators via twists of $r_{\rm DJ}$\,. 
One may also consider the extension of the twisted $R$ operators 
by adding bilinear terms of fermionic root generators. 
It should be noted that the latter cannot be obtained with the twists. 
Thus there are the two classes: 
1) Jordanian $R$ operators stemming from twists and 
2) extended Jordanian $R$ operators.   
We will introduce some examples below.

\subsubsection*{1) Jordanian $R$ operators from twists} 

The first example is twists by simple roots.    
Then the corresponding subsectors of the superstring action are deformed.  
For instance, let us consider twists by positive simple root generators 
$E_{k,k+1}$ ($k=1,\ldots, \check{4} , \ldots,7$)\footnote{
For $k=4$\,, the simple root $E_{45}=\bar S_{45}$ is fermionic and it is regarded as a fermionic twist.}.     
Then the associated classical $r$-matrix is given by  
\begin{align}
r_{\rm tw}^{(k,k+1)} = [\Delta(E_{k,k+1}), r_{\rm DJ}]=c E_{k,k+1}\wedge 
\left(E_{kk}(-1)^{\bar k}-E_{k+1,k+1}(-1)^{\overline{k+1}}\right)\,.  
\label{simpleR}
\end{align}
The twists give rise to deformations of the AdS$_3$ or S$^3$ subspace. 
For each of the values $k=1,2,3$\,, the resulting geometry is given by a deformed AdS$_3$ spacetime.  
It would contain a three-dimensional Schr\"odinger spacetime 
and may be regarded as a generalization of the previous works 
\cite{KY-Sch, Jordanian-KMY}. The explicit relation will be presented in \cite{future}. 

\medskip 

More interesting examples are deformations of either AdS$_5$ or S$^5$. 
These partial deformations are realized by twists with the 
maximal bosonic generators $E_{14}=P_{14}$ in $\alg{su}(2,2)$ 
and $E_{58}=R_{58}$ in $\alg{su}(4)$\,, respectively 
\footnote{For the map between the $E_{ij}$ generators 
and the superconformal generators, see Appendix \ref{app:sconf}.};
\begin{align}
\text{AdS}_5 :& \quad 
r_{\rm tw}^{(1,4)} = [\Delta(E_{14}), r_{\rm DJ}]= c\Bigl(E_{14}\wedge (E_{11}-E_{44}) 
-2 \sum_{\kappa=2,3}E_{1\kappa}\wedge E_{\kappa 4} \Bigr)\,,
\label{ads5R}
\\
\text{S}^5 :& \quad
r_{\rm tw}^{(5,8)} = [\Delta(E_{58}), r_{\rm DJ}]= c\Bigl(E_{58}\wedge (-E_{55}+E_{88}) 
+2 \sum_{k=6,7}E_{5k}\wedge E_{k 8} \Bigr)\,.
\label{s5R}
\end{align}
The deformation of S$^5$ should be interesting because 
it would provide a simpler geometry without deforming AdS$_5$\,. 
The associated linear operator acts on the generators as follows: 
\begin{align}
&R_{\rm tw}^{(5,8)} (E_{55}) = +cE_{58} \,,
&&R_{\rm tw}^{(5,8)} (E_{k5}) = +2cE_{k8}\,,\no \\
&R_{\rm tw}^{(5,8)} (E_{88}) = -cE_{58} \,,
&&R_{\rm tw}^{(5,8)} (E_{8k}) = -2cE_{5k}\,, \no \\
&R_{\rm tw}^{(5,8)} (E_{85}) = c(-E_{55}+E_{88}) \,, 
&&  R_{\rm tw}^{(5,8)} (\text{others}) =0\,, \no 
\end{align}
where $k=6,7$\,. 

\paragraph{Remarks}
More generally, the Reshetikhin twist \cite{R} or the Jordanian twist \cite{Jordanian,KLM} 
is closely related to the present 
prescription. The Jordanian twists for Lie superalgebras are considered in 
\cite{Tolstoy, BLT, ACS, ACCYZ}. 
The relation will be elaborated somewhere else. 

\medskip 

As a side remark, we have worked with a particular choice of the simple 
roots associated with the Dynkin diagram O-O-O-X-O-O-O of the superconformal algebra. 
It would be also interesting to see the twisting based on the different choice 
of simple roots such as O-X-O-O-O-X-O.

\subsubsection*{2) Extended Jordanian $R$ operators} 

Let us now consider some extensions of the twisted classical $r$-matrices 
given in \eqref{simpleR}, \eqref{ads5R} and \eqref{s5R}. 
Recall that these are obtained by twisting $r_{\rm DJ}$\,. Here we are concerned with 
some extensions of the twisted $r$-matrices, which are not described as twists. 

\medskip 

It is easy to see that a linear combination of \eqref{ads5R} and \eqref{s5R}
\begin{align}
r_{\rm tw}^{(1,4), (5,8)} \equiv c_1 r_{\rm tw}^{(1,4)} +c_2 r_{\rm tw}^{(5,8)}
\end{align}
with $c=1$ is also a solution of CYBE, due to the relation  
\begin{align}
[r_{\rm tw}^{(1,4)}, r_{\rm tw}^{(5,8)}]=0\,. 
\end{align}
The $r$-matrix $r_{\rm tw}^{(1,4), (5,8)}$ implies independent deformations of AdS$^5$ and S$^5$ 
with different parameters $c_1$ and $c_2$ respectively. 

\medskip 

Furthermore, these $r$-matrices may be extended to contain supercharges in their tails, 
including two parameters, like 
\begin{align}
&\wt{r}_{\rm tw}^{(1,4)} = E_{14}\wedge (\al E_{11}-\be E_{44}) 
-(\al+\be) \sum_{j\neq 1,4}E_{1j}\wedge E_{j 4} \,,
\\
&\wt{r}_{\rm tw}^{(5,8)} = E_{58}\wedge (\al' E_{55}-\be' E_{88}) 
-(\al'+\be') \sum_{j\neq 5,8}E_{5j}\wedge E_{j 8}\,.
\end{align}
Here $\al,\be, \al', \be'$ are arbitrary parameters. 
The extended $r$-matrices satisfy CYBE \eqref{CYB}. 

\medskip 

As a remark, it would not be obvious that the multi-parameter 
deformations may lead to consistent string theories. 
For the vanishing $\beta$-function on the world-sheet, 
there may be additional constraints on the deformation parameters.

\subsubsection*{Comments on fermionic twists} 

One may think of twists by fermionic generators.  
However, it seems that the resulting $r$-matrices
do not satisfy CYBE \eqref{CYB}, in general. 

\medskip 

As an example, let us consider $E_{45}=\bar S_{45}$.  
This is a simple root generator but it gives rise to the maximal twist. 
That is, the corresponding geometry is also maximally deformed. 
The associated classical $r$-matrix is given by 
\begin{align}
r_{\rm tw}^{(4,5)} = [\Delta(E_{45}), r_{\rm DJ}] 
= c \Bigl[(E_{44}+E_{55})\wedge E_{45} 
+ 2\sum_{\kappa=1}^3 E_{4\kappa} \wedge E_{\kappa 5} 
- 2\sum_{k=6}^8 E_{4k} \wedge E_{k5} \Bigr]\,. 
\label{twR12}
\end{align}
Note that $c$ is a Grassmann odd element \cite{Tolstoy}, 
so that the $r$-matrix is Grassmann even. 
However it does not seem to be a solution of CYBE. 

\medskip

The only exception is obtain by the maximal root, (Also see Appendix B.2)  
\begin{align}
r_{\rm tw}^{(1,8)}=[\Delta(E_{18}), r_{\rm DJ}]=-c E_{18} \wedge (E_{11}+E_{88})\,. 
\label{fer18}
\end{align}
This is a solution of CYBE. 
For this fermionic twist, we have no clear understanding for the physical interpretation 
because the deformation is measured by a Grassmann odd parameter. It would be interesting to 
interpret the fermionic twist in type IIB supergravity.

\subsection{Jordanian deformed action}

The next is to consider Jordanian deformations of the 
classical action of the AdS$_5\times$S$^5$ superstring. 
Although the construction is almost parallel to the one in \cite{DMV2}, it is necessary to take account of 
small modifications coming from the fact that the Jordanian linear operator $R_{\rm Jor}$ 
satisfies CYBE rather than mCYBE. 
\medskip 

In the following, $R_{\rm Jor}$ is used as a representative of 
arbitrary (extended) Jordanian $R$ operators\footnote{ 
The (extended) Jordanian operators are easily derived from the tensorial $r$-matrix presented in Sec.\,3.1 
by using the relations \eqref{RR12} and the inner product \eqref{inp}\,.}.  
The detail expression of $R_{\rm Jor}$ is not relevant to the subsequent analysis.  

\medskip

The Jordanian deformed classical action is given by 
\begin{eqnarray}
S = -\frac{1}{2} \int^{\infty}_{-\infty}\!\!d\tau\int^{2\pi}_0\!\!d\sigma\,P_-^{\alpha\beta}\, 
{\rm Str}\left(A_{\alpha}\, d\circ \frac{1}{1-\eta 
\left[R_{\rm Jor}\right]_g 
\circ d} (A_{\beta}) \right)\,. 
\label{action}
\end{eqnarray}
Here, by using Jordanian $R$-matrix 
$R_{\rm Jor}$\,,
a chain of the operations 
$\left[R_{\rm Jor}\right]_g$ is defined as 
\begin{eqnarray}
\left[R_{\rm Jor}\right]_g(M) 
\equiv Ad_g^{-1} \circ R_{\rm Jor}  \circ Ad_g(M) = g^{-1}R_{\rm Jor}(g M g^{-1})g\,.
\end{eqnarray}
In the present case, $d$ and $\tilde{d}$ are not deformed and do not contain $\eta$ like 
\begin{eqnarray}
d \equiv P_1 + 2 P_2 -P_3\,, \qquad \tilde{d} \equiv -P_1 + 2 P_2 + P_3\,, 
\end{eqnarray}
and the overall factor of the action (\ref{q-action}) is not needed to be multiplied. 
As in the case of \cite{DMV2},  
the equations of motion can be written simply 
with the following quantities: 
\begin{eqnarray}
J_{\alpha} \equiv \frac{1}{1-\eta \left[R_{\rm Jor}\right]_g \circ d}\,(A_{\alpha})\,, 
\qquad J_-^{\alpha} \equiv P_-^{\alpha\beta}J_{\beta}\,, \\
\widetilde{J}_{\alpha} \equiv \frac{1}{1+\eta \left[R_{\rm Jor}\right]_g \circ \tilde{d}}\,(A_{\alpha})\,, 
\qquad \widetilde{J}_+^{\alpha} \equiv P_+^{\alpha\beta}\widetilde{J}_{\beta}\,. \nonumber
\end{eqnarray}

\medskip 

There are two ways to rewrite the action given in (\ref{action})\,. 
The first is based on $J_\alpha$ and the action is written as  
\begin{eqnarray}
S&=&-\frac{1}{4}
\int^{\infty}_{-\infty}\!\!d\tau\int^{2\pi}_0\!\!d\sigma\,
\left(\gamma^{\alpha\beta}-\epsilon^{\alpha\beta}\right){\rm Str}\left(J_\alpha d\left(J_\beta\right)\right) \nonumber \\
&&+\frac{\eta}{4}
\int^{\infty}_{-\infty}\!\!d\tau\int^{2\pi}_0\!\!d\sigma\,
\left(\gamma^{\alpha\beta}-\epsilon^{\alpha\beta}\right){\rm Str}\left(\left[R_{\rm Jor}\right]_g\circ d(J_\alpha) d\left(J_\beta\right)\right) \nonumber \\
&=&-\frac{1}{2}
\int^{\infty}_{-\infty}\!\!d\tau\int^{2\pi}_0\!\!d\sigma\,
\gamma^{\alpha\beta}{\rm Str}\left(J^{(2)}_\alpha J^{(2)}_\beta\right)
-\frac{1}{2}
\int^{\infty}_{-\infty}\!\!d\tau\int^{2\pi}_0\!\!d\sigma\,
\epsilon^{\alpha\beta}{\rm Str}\left(J^{(1)}_\alpha J^{(3)}_\beta\right) \nonumber \\
&&+\frac{\eta}{4}
\int^{\infty}_{-\infty}\!\!d\tau\int^{2\pi}_0\!\!d\sigma\,
\epsilon^{\alpha\beta}{\rm Str}\left(d(J_\alpha)\left[R_{\rm Jor}\right]_g\circ d\left(J_\beta\right)\right)\,. \label{action-J}
\end{eqnarray}
The second is based on $\widetilde{J}_\alpha$ and the action becomes 
\begin{eqnarray}
S&=&-\frac{1}{4}
\int^{\infty}_{-\infty}\!\!d\tau\int^{2\pi}_0\!\!d\sigma\,
\left(\gamma^{\alpha\beta}-\epsilon^{\alpha\beta}\right){\rm Str}\left(\widetilde{d}(\widetilde{J}_\alpha)\widetilde{J}_\beta\right) \nonumber \\
&&-\frac{\eta}{4}
\int^{\infty}_{-\infty}\!\!d\tau\int^{2\pi}_0\!\!d\sigma\,
\left(\gamma^{\alpha\beta}-\epsilon^{\alpha\beta}\right){\rm Str}\left(\widetilde{d}(\widetilde{J}_\alpha)\left[R_{\rm Jor}\right]_g\circ\widetilde{d}(\widetilde{J}_\beta)\right) \nonumber \\
&=&-\frac{1}{2}
\int^{\infty}_{-\infty}\!\!d\tau\int^{2\pi}_0\!\!d\sigma\,
\gamma^{\alpha\beta}{\rm Str}\left(\widetilde{J}^{(2)}_\alpha\widetilde{J}^{(2)}_\beta\right)
-\frac{1}{2}
\int^{\infty}_{-\infty}\!\!d\tau\int^{2\pi}_0\!\!d\sigma\,
\epsilon^{\alpha\beta}{\rm Str}\left(\widetilde{J}^{(1)}_\alpha\widetilde{J}^{(3)}_\beta\right) \nonumber \\
&&+\frac{\eta}{4}
\int^{\infty}_{-\infty}\!\!d\tau\int^{2\pi}_0\!\!d\sigma\,
\epsilon^{\alpha\beta}{\rm Str}\left(\widetilde{d}(\widetilde{J}_\alpha)\left[R_{\rm Jor}\right]_g\circ\widetilde{d}(\widetilde{J}_\beta)\right)\,. \label{action-tildeJ}
\end{eqnarray}
The two expressions are useful to discuss the Virasoro conditions and the $\kappa$-invariance. 

\medskip 

Then equations of motion are given by 
\begin{eqnarray}
\mathcal{E} = d(\partial_{\alpha} J_-^{\alpha}) + \tilde{d}(\partial_{\alpha}\widetilde{J}_+^{\alpha}) 
+ [\widetilde{J}_{+\alpha}, d (J_-^{\alpha})] + [J_{-\alpha},\tilde{d}(\widetilde{J}_+^{\alpha})] = 0\,, 
\label{eom-null}
\end{eqnarray}
and the flatness condition is represented by 
\begin{eqnarray}
\mathcal{Z} &=& \frac{1}{2}\epsilon^{\alpha\beta}\left(
\partial_{\alpha}A_{\beta} - \partial_{\beta}A_{\alpha} + [A_{\alpha},A_{\beta}]
\right) \nonumber \\ 
&=&\partial_{\alpha} \widetilde{J}_+^{\alpha} - \partial_{\alpha}J_-^{\alpha} 
+ [J_{-\alpha}, \widetilde{J}_+^{\alpha}] + \eta \left[R_{\rm Jor}\right]_g(\mathcal{E}) = 0\,. \label{flat-null}
\end{eqnarray}
Note that the flatness condition does not contain the $\eta^2$ terms, 
in comparison to the one given in (\ref{q-flat})\,. This modification comes from the fact 
that the Jordanian operator $R_{\rm Jor}$ 
satisfies CYBE, rather than mCYBE. 

\medskip 

For later computations, it is convenient to decompose the equations of motion \eqref{eom-null} 
and the flatness condition \eqref{flat-null} as follows: 
\begin{eqnarray}
&&\partial_\alpha\widetilde{J}^{\alpha(0)}_+\!-\partial_\alpha J^{\alpha(0)}_-\!
+\!\left[J^{(0)}_{-\alpha},\widetilde{J}^{\alpha(0)}_+\right]\!
+\!\left[J^{(1)}_{-\alpha},\widetilde{J}^{\alpha(3)}_+\right]\!
+\!\left[J^{(2)}_{-\alpha},\widetilde{J}^{\alpha(2)}_+\right]\!
+\!\left[J^{(3)}_{-\alpha},\widetilde{J}^{\alpha(1)}_+\right]=0\,, 
\label{eom2} \\
&&\left[J^{(3)}_{-\alpha},\widetilde{J}^{\alpha(2)}_+\right]=0\,, \nonumber \\
&&\partial_\alpha\widetilde{J}^{\alpha(1)}_+-\partial_\alpha J^{\alpha(1)}_-
+\left[J^{(0)}_{-\alpha},\widetilde{J}^{\alpha(1)}_+\right]
+\left[J^{(1)}_{-\alpha},\widetilde{J}^{\alpha(0)}_+\right]
+\left[J^{(2)}_{-\alpha},\widetilde{J}^{\alpha(3)}_+\right]=0\,, \nonumber \\
&&\partial_\alpha\widetilde{J}^{\alpha(2)}_+
+\left[J^{(0)}_{-\alpha},\widetilde{J}^{\alpha(2)}_+\right]
+\left[J^{(3)}_{-\alpha},\widetilde{J}^{\alpha(3)}_+\right]=0\,, \nonumber \\
&&\partial_\alpha J^{\alpha(2)}_-
-\left[J^{(1)}_{-\alpha},\widetilde{J}^{\alpha(1)}_+\right]
-\left[J^{(2)}_{-\alpha},\widetilde{J}^{\alpha(0)}_+\right]=0\,, \nonumber \\
&&\left[J^{(2)}_{-\alpha},\widetilde{J}^{\alpha(1)}_+\right]=0\,, \nonumber \\
&&\partial_\alpha\widetilde{J}^{\alpha(3)}_+-\partial_\alpha J^{\alpha(3)}_-
+\left[J^{(0)}_{-\alpha},\widetilde{J}^{\alpha(3)}_+\right]
+\left[J^{(1)}_{-\alpha},\widetilde{J}^{\alpha(2)}_+\right]
+\left[J^{(3)}_{-\alpha},\widetilde{J}^{\alpha(0)}_+\right]=0\,. \nonumber
\end{eqnarray}
Then the Lax pair is given by 
\begin{eqnarray}
&&M_-^{\alpha} = J^{\alpha(0)}_-+\lambda J^{\alpha(1)}_-+\lambda^2 J^{\alpha(2)}_-+\lambda^{-1}J^{\alpha(3)}_-\,, \\
&&L_{+}^{\alpha} = \widetilde{J}^{\alpha(0)}_++\lambda\widetilde{J}^{\alpha(1)}_++\lambda^{-2}\widetilde{J}^{\alpha(2)}_++\lambda^{-1}\widetilde{J}^{\alpha(3)}_+\,.
\end{eqnarray}
Note that the $\eta^2$ terms are not contained again, 
in comparison to the Lax pair given in (\ref{q-Lax})\,, 
while the parameter $\eta$ is still contained in 
$J_-^{\alpha(n)}$ and $\widetilde{J}_+^{\alpha(n)}$ $(n=0,\ldots,3)$\,. 
With $\mathcal{L}_{\alpha} \equiv L_{+\alpha} + M_{-\alpha}$\,, it is an easy task to show that the zero curvature condition (\ref{zero}) 
is equivalent to the equation of motion (\ref{eom-null}) and the flatness condition (\ref{flat-null})\,.  

\medskip 

The next is to consider the Virasoro conditions.  
The expression given in \eqref{action-J} leads to the Virasoro conditions,  
\begin{eqnarray}
{\rm Str}\left(J^{(2)}_\alpha J^{(2)}_\beta\right)-\frac{1}{2}\gamma_{\alpha\beta}\gamma^{\rho\sigma}{\rm Str}\left(J^{(2)}_\rho J^{(2)}_\sigma\right)=0\,. 
\label{Virasoro1}
\end{eqnarray}
On the other hand, the expression in \eqref{action-tildeJ} gives rise to  
\begin{eqnarray}
{\rm Str}\left(\widetilde{J}^{(2)}_\alpha\widetilde{J}^{(2)}_\beta\right)-\frac{1}{2}\gamma_{\alpha\beta}\gamma^{\rho\sigma}{\rm Str}\left(\widetilde{J}^{(2)}_\rho\widetilde{J}^{(2)}_\sigma\right)=0\,. 
\label{Virasoro2}
\end{eqnarray}
The above two representations of the Virasoro conditions given in \eqref{Virasoro1} and \eqref{Virasoro2} 
should be equivalent. 

\subsection{$\kappa$-symmetry}

Let us consider the $\kappa$-symmetry of the action (\ref{action})\,. 

\medskip 

We consider a fermionic local transformation (called the $\kappa$-transformation) of $g$ given by  
\begin{eqnarray}
\delta g=g\epsilon\,, \qquad
\epsilon\equiv\left(1-\eta \left[R_{\rm Jor}\right]_g\right)\rho^{(1)}
+\left(1+\eta \left[R_{\rm Jor}\right]_g\right)\rho^{(3)}\,, 
\label{kappa1}
\end{eqnarray}
where $\rho^{(1)}$ and $\rho^{(3)}$ are arbitrary functions on the string world-sheet 
to be determined later, and hence $\epsilon$ also depends on the world-sheet coordinates. 
Then the variation of the action given in \eqref{action} is described as  
\begin{eqnarray}
\delta_g S&=&\frac{1}{2}
\int^{\infty}_{-\infty}\!\!d\tau\!\int^{2\pi}_0\!\!d\sigma\, {\rm Str}\left(\epsilon\,{\mathcal E}\right) \\
&=&\frac{1}{2}
\int^{\infty}_{-\infty}\!\!d\tau\int^{2\pi}_0\!\!d\sigma\, {\rm Str}\left(
\rho^{(1)}P_3\circ\left(1+\eta \left[R_{\rm Jor}\right]_g\right)({\mathcal E}) \right. \nonumber \\
&&\hspace{5cm}\left.+
\rho^{(3)}P_1\circ\left(1-\eta \left[R_{\rm Jor}\right]_g\right)({\mathcal E})\right) \nonumber \\
&=&-2
\int^{\infty}_{-\infty}\!\!d\tau\int^{2\pi}_0\!\!d\sigma\, {\rm Str}\left(\rho^{(1)}\left[J^{(2)}_{-\alpha},\widetilde{J}^{\alpha(1)}_+\right]+
\rho^{(3)}\left[\widetilde{J}^{(2)}_{+\alpha},J^{\alpha(3)}_-\right]\right)\,.  \nonumber
\end{eqnarray}
Here the following relations have been used in the second equality, 
\begin{eqnarray}
P_1\circ\left(1-\eta \left[R_{\rm Jor}\right]_g\right)({\mathcal E})
=-4\left[\widetilde{J}^{(2)}_{+\alpha},J^{\alpha(3)}_-\right]-P_1({\mathcal Z})\,, \\
P_3\circ\left(1+\eta \left[R_{\rm Jor}\right]_g\right)({\mathcal E})
=-4\left[J^{(2)}_{-\alpha},\widetilde{J}^{\alpha(1)}_+\right]+P_3({\mathcal Z})\,. \nonumber
\end{eqnarray}

\medskip 

Now let the forms of $\rho^{(1)}$ and $\rho^{(3)}$ be   
\begin{eqnarray}
\rho^{(1)}=i\kappa^{\alpha(1)}_+J^{(2)}_{-\alpha}+J^{(2)}_{-\alpha}i\kappa^{\alpha(1)}_+\,, \qquad
\rho^{(3)}=i\kappa^{\alpha(3)}_-\widetilde{J}^{(2)}_{+\alpha}+\widetilde{J}^{(2)}_{+\alpha}i\kappa^{\alpha(3)}_-\,. 
\end{eqnarray}
Note that these forms are compatible to the grading assignment. 
Then one can show the relation
\begin{eqnarray}
{\rm Str}\left(\rho^{(1)}\left[J^{(2)}_{-\alpha},\widetilde{J}^{\alpha(1)}_+\right]\right) 
= {\rm Str}\left(J^{(2)}_{-\alpha}J^{(2)}_{-\beta}\left[\widetilde{J}^{\alpha(1)}_+,i\kappa^{\beta(1)}_+\right]\right)\,. 
\end{eqnarray}
The derivation is the following,  
\begin{eqnarray}
{\rm Str}\left(\rho^{(1)}\left[J^{(2)}_{-\alpha},\widetilde{J}^{\alpha(1)}_+\right]\right) 
&=&{\rm Str}\left[\left(i\kappa^{\tau(1)}_+J^{(2)}_{-\tau}+J^{(2)}_{-\tau}i\kappa^{\tau(1)}_+
+i\kappa^{\sigma(1)}_+J^{(2)}_{-\sigma}+J^{(2)}_{-\sigma}i\kappa^{\sigma(1)}_+\right)\right. \nonumber \\
&&\hspace{1cm}\left.\times\left(J^{(2)}_{-\tau}\widetilde{J}^{\tau(1)}_+-\widetilde{J}^{\tau(1)}_+J^{(2)}_{-\tau}
+J^{(2)}_{-\sigma}\widetilde{J}^{\sigma(1)}_+-\widetilde{J}^{\sigma(1)}_+J^{(2)}_{-\sigma}\right)\right] \nonumber \\
&=&{\rm Str}\left[J^{(2)}_{-\tau}J^{(2)}_{-\tau}\left(\widetilde{J}^{\tau(1)}_+i\kappa^{\tau(1)}_+-i\kappa^{\tau(1)}_+\widetilde{J}^{\tau(1)}_+\right)\right. \nonumber \\
&&\hspace{1cm}+J^{(2)}_{-\tau}J^{(2)}_{-\sigma}\left(\widetilde{J}^{\tau(1)}_+i\kappa^{\sigma(1)}_+-i\kappa^{\sigma(1)}_+\widetilde{J}^{\tau(1)}_+\right) \nonumber \\
&&\hspace{1cm}+J^{(2)}_{-\sigma}J^{(2)}_{-\tau}\left(\widetilde{J}^{\sigma(1)}_+i\kappa^{\tau(1)}_+-i\kappa^{\tau(1)}_+\widetilde{J}^{\sigma(1)}_+\right) \nonumber \\
&&\hspace{1cm}+\left.J^{(2)}_{-\sigma}J^{(2)}_{-\sigma}\left(\widetilde{J}^{\sigma(1)}_+i\kappa^{\sigma(1)}_+-i\kappa^{\sigma(1)}_+\widetilde{J}^{\sigma(1)}_+\right)\right] \nonumber \\
&=&{\rm Str}\left(J^{(2)}_{-\alpha}J^{(2)}_{-\beta}\left[\widetilde{J}^{\alpha(1)}_+,i\kappa^{\beta(1)}_+\right]\right)\,. \nonumber
\end{eqnarray}
The second equality comes from the fact that $J^{(2)}_{-\tau}$ is proportional to $J^{(2)}_{-\sigma}$\,. 
Similarly, one can show the relation, 
\begin{eqnarray}
{\rm Str}\left(\rho^{(3)}\left[\widetilde{J}^{(2)}_{+\alpha},J^{\alpha(3)}_-\right]\right)
={\rm Str}\left(\widetilde{J}^{(2)}_{+\alpha}\widetilde{J}^{(2)}_{+\beta}\left[J^{\alpha(3)}_-,i\kappa^{\beta(3)}_-\right]\right)\,. 
\end{eqnarray}
Furthermore, for any grade $2$ traceless matrix $A^{(2)}_{\pm\alpha}$\,, 
the following relation is satisfied \cite{AF-review}\,, 
\begin{eqnarray}
A^{(2)}_{\pm\alpha}A^{(2)}_{\pm\beta}=\frac{1}{8}{\rm Str}\left(A^{(2)}_{\pm\alpha}A^{(2)}_{\pm\beta}\right)\Upsilon+c_{\alpha\beta}{\bf 1}_8\,, 
\end{eqnarray}
by using the matrix representation, 
where $\Upsilon$ is the following $8\times 8$ matrix: 
\begin{eqnarray}
\Upsilon={\rm diag}({\bf 1}_4,-{\bf 1}_4)\,. 
\end{eqnarray}
Thus the following relations are obtained, 
\begin{eqnarray}
{\rm Str}\left(\rho^{(1)}\left[J^{(2)}_{-\alpha},\widetilde{J}^{\alpha(1)}_+\right]\right)
&=&\frac{1}{8}{\rm Str}\left(J^{(2)}_{-\alpha}J^{(2)}_{-\beta}\right)
{\rm Str}\left(\Upsilon\left[\widetilde{J}^{\alpha(1)}_+,i\kappa^{\beta(1)}_+\right]\right)\,, \label{rel1}\\
{\rm Str}\left(\rho^{(3)}\left[\widetilde{J}^{(2)}_{+\alpha},J^{\alpha(3)}_-\right]\right)
&=&\frac{1}{8}{\rm Str}\left(\widetilde{J}^{(2)}_{+\alpha}\widetilde{J}^{(2)}_{+\beta}\right)
{\rm Str}\left(\Upsilon\left[J^{\alpha(3)}_-,i\kappa^{\beta(3)}_-\right]\right)\,. \label{rel2}
\end{eqnarray}
With the relations (\ref{rel1}) and (\ref{rel2})\,, the variation of the classical action 
\eqref{action} under the transformation \eqref{kappa1} is evaluated as  
\begin{eqnarray}
\delta_g S&=&-\frac{1}{4}
\int^{\infty}_{-\infty}\!\!d\tau\int^{2\pi}_0\!\!d\sigma\, {\rm Str}\left({\rm Str}\left(J^{(2)}_{-\alpha}J^{(2)}_{-\beta}\right)
\Upsilon\left[\widetilde{J}^{\alpha(1)}_+,i\kappa^{\beta(1)}_+\right] 
\right. \nonumber \\ &&\hspace{4cm}\left.
+
{\rm Str}\left(\widetilde{J}^{(2)}_{+\alpha}\widetilde{J}^{(2)}_{+\beta}\right)
\Upsilon\left[J^{\alpha(3)}_-,i\kappa^{\beta(3)}_-\right]\right)\,. 
\end{eqnarray}

\medskip 

Then we will show that this variation is canceled out with the variation of the action 
with respect to the world-sheet metric $\gamma^{\alpha\beta}$\,. 
Let the variation of $\gamma^{\alpha\beta}$ be  
\begin{eqnarray}
\delta \gamma^{\alpha\beta}&=&-\frac{1}{4}{\rm Str}\left(
\Upsilon\left[\widetilde{J}^{\alpha(1)}_+,i\kappa^{\beta(1)}_+\right]
+\Upsilon\left[\widetilde{J}^{\beta(1)}_+,i\kappa^{\alpha(1)}_+\right] \right. \\
&&\hspace{2cm}\left. +\Upsilon\left[J^{\alpha(3)}_-,i\kappa^{\beta(3)}_-\right]
+\Upsilon\left[J^{\beta(3)}_-,i\kappa^{\alpha(3)}_-\right]\right)\,. \nonumber
\end{eqnarray}
Then, by using the expressions of the classical action given in \eqref{action-J} and \eqref{action-tildeJ}\,, 
the variation of the action is evaluated as 
\begin{eqnarray}
\delta_\gamma S&=&\frac{1}{4}
\int^{\infty}_{-\infty}\!\!d\tau\int^{2\pi}_0\!\!d\sigma\,\left[
{\rm Str}\left(\Upsilon\left[\widetilde{J}^{\alpha(1)}_+,i\kappa^{\beta(1)}_+\right]\right)
{\rm Str}\left(J^{(2)}_\alpha J^{(2)}_\beta\right) \right.\nonumber \\
&&\hspace{3.5cm}\left.+{\rm Str}\left(\Upsilon\left[J^{\alpha(3)}_-,i\kappa^{\beta(3)}_-\right]\right) 
{\rm Str}\left(\widetilde{J}^{(2)}_\alpha\widetilde{J}^{(2)}_\beta\right)\right] \nonumber \\
&=&\frac{1}{4}
\int^{\infty}_{-\infty}\!\!d\tau\int^{2\pi}_0\!\!d\sigma\,\left[
{\rm Str}\left(\Upsilon\left[\widetilde{J}^{\alpha(1)}_+,i\kappa^{\beta(1)}_+\right]\right)
{\rm Str}\left(J^{(2)}_{-\alpha}J^{(2)}_{-\beta}\right) \right.\nonumber \\
&&\hspace{3.5cm}\left.+{\rm Str}\left(\Upsilon\left[J^{\alpha(3)}_-,i\kappa^{\beta(3)}_-\right]\right) 
{\rm Str}\left(\widetilde{J}^{(2)}_{+\alpha}\widetilde{J}^{(2)}_{+\beta}\right)\right]\,. 
\end{eqnarray}
In order to show the second equality, the following relations have been used,  
\begin{eqnarray}
A^\alpha_\pm B_\alpha=A^\alpha_\pm B_{\pm\alpha}+A^\alpha_\pm B_{\mp\alpha}=A^\alpha_\pm B_{\mp\alpha}\,. 
\end{eqnarray}
%The second equality follows from the fact that 
%inner products between $A^\alpha_+$ and $B^\alpha_-$ are identically zero. 
Thus, the total variation of the classical action \eqref{action} becomes zero,  
\begin{eqnarray}
\delta_g S+\delta_\gamma S=0\,. 
\end{eqnarray}
That is, the action \eqref{action} is invariant under the $\kappa$-transformation.

\section{Conclusion and Discussion}

We have discussed Jordanian deformations of the AdS$_5\times$S$^5$ superstring action. 
The description to construct Jordanian $R$ operators via twists 
has been explained in detail. Notably the Jordanian
$R$ operators satisfy CYBE rather than mCYBE
and they have non-vanishing Cartan charge.    
Then we have constructed the Jordanian deformed action 
that preserves the $\kappa$-symmetry. The Lax pair has also been presented.

\medskip 

It should be remarked that partial deformations are possible in our procedure. 
This fact implies that one may perform a deformation only for the S$^5$ part, 
for example. Then the background geometry would be much simpler because the AdS$_5$ part is 
not modified and the gauge-theory dual would be identified with a deformation of 
the scalar sector such as Leigh-Strassler deformations \cite{LS}. 
A promising way is to consider a twist of the $q$-deformation of the $SO(6)$ sector 
argued in \cite{B-Cherkis,B-Correa}. 
As a matter of course, even for the maximal twist, the metric of the twisted geometry can  
be determined, for example, by following \cite{ABF}. 
For the background geometries, we will report on the result 
in the near future \cite{future}. 

\medskip

In principle, it should be possible to classify all of the skew-symmetric 
classical $r$-matrices of $\alg{gl}(4|4)$ and its real form $\alg{su}(2,2|4)$. 
The classification enables us to reveal all of the possible deformations of 
the AdS$_5\times$S$^5$ superstring from the algebraic point of view.  

\medskip 

We believe that the study of integrable deformations of the AdS$_5\times$S$^5$ superstring 
would shed light on new aspects of the integrable structure behind the AdS/CFT correspondence.

\subsection*{Acknowledgments}

We would like to thank T.~Kameyama, M.~Magro, S.~Moriyama and B.~Vicedo for useful discussions. 
The work of IK was supported by the Japan Society for the Promotion of Science (JSPS). 
T.M. also thanks G.~Arutyunov for his comments on the first version on arXiv 
and R.~Borsato for explaining his recent work \cite{ABF}.  
T.M. is supported by the Netherlands Organization for Scientific 
Research (NWO) under the VICI grant 680-47-602.  
T.M.'s work is also part of the ERC Advanced grant research programme 
No. 246974, ``Supersymmetry: a window to non-perturbative physics" 
and of the D-ITP consortium, a program of the NWO that is funded by the 
Dutch Ministry of Education, Culture and Science (OCW).

\appendix

\section*{Appendix}

\section{Notations of superconformal generators \label{app:sconf}}

In this paper, we work with the $\alg{gl}(4|4)$ generators rather than 
$\alg{u}(2,2|4)$ generators because the former generators are more convenient for the algebraic argument. 
The superconformal algebra $\alg{u}(2,2|4)$ is obtained from $\alg{gl}(4|4)$ 
by imposing a suitable condition.  
Thus, we will spell out the explicit relations among the generators. 
This is enough for our purpose. 

\medskip

The Lie superalgebra $\alg{gl}(4|4)$ is a $32|32$ dimensional algebra and 
generated by $E_{ij}$ with $i,j=1,\cdots, 8$ satisfying the relations\footnote{
The commutator is assumed to be supercommutator here and 
also in \eqref{mCYB}, \eqref{CYB}, \eqref{gl44} and \eqref{glnm}\,. 
The other commutators are not graded in constructing the action of the AdS$_5\times$S$^5$ superstring, 
because we consider a Grassmann envelope of the superalgebra by following \cite{DMV2}.
}, 
\begin{align}
[E_{ij},E_{kl}]=\de_{kj} E_{il}-\de_{il} E_{kj} (-1)^{(\bar i+\bar j)(\bar k+\bar l)}\,. 
\label{gl44}
\end{align}
Here the parity of indices are defined as 
~$\bar i=0$~~for $i=1, \cdots , 4$ and 
~$\bar i=1$~~for $i=5, \cdots , 8$. 

\medskip

The invariant super-symmetric non-degenerate linear form is defined as   
\begin{align}
\lan E_{ij} , E_{kl} \ran =\de_{kj} \de_{il} (-1)^{\bar j}\,, 
\label{inp}
\end{align}
with $i,j,k,l=1,\cdots, 8$\,, which satisfies the following properties
\begin{align}
&\lan E_{ij} , E_{kl} \ran = \lan E_{kl} , E_{ij} \ran(-1)^{(\bar i+\bar j)(\bar k+\bar l)}\,, \no \\
&\lan E_{ij} , E_{kl} \ran = 0 \qquad \quad  \text{for} \quad \bar i+\bar j\neq \bar k+\bar l\,.  
\end{align}

\medskip

The bosonic part of the superconformal algebra is related to $\alg{gl}(4|4)$ generators as 
\begin{align}
&L_{\al\be}=E_{\al\be}-\tfrac{1}{2}\de_{\al\be} E_{\la\la}\,, 
&& D=\tfrac{1}{2} (E_{\la\la} -E_{\dot \la \dot \la})\,,   
&& P_{\al\dot \be}=E_{\al\dot \be}\,, \no\\ 
&\bar L_{\dot \al\dot \be}=E_{\dot \al\dot \be}-\tfrac{1}{2}\de_{\dot \al\dot \be} E_{\dot \la\dot \la}\,, 
&& C=\tfrac{1}{2} (E_{\la\la} +E_{\dot \la \dot \la}+E_{ll})\,,   
&& K_{\dot \al \be}=E_{\dot \al \be}\,, \no\\ 
&R_{ab}=E_{ab}-\tfrac{1}{4}\de_{ab} E_{ll}\,, 
&& B= -\tfrac{1}{2} E_{ll}\,,  
\end{align}
where $\al,\be,\la=1,2$, $\dot \al,\dot \be, \dot \la=3,4$ and 
$a, b, l= 5, \cdots, 8$\,.  
The conformal algebra $\alg{su}(2,2)$ contains two $\alg{su}(2)$ subalgebras 
generated by $L_{\al\be}$ and $\bar L_{\dot \al\dot \be}$ as well as  
the translations $P_{\al\dot \be}$ and the conformal boosts $K_{\dot \al \be}$\,. 
The R-symmetry $\alg{su}(4)$ is generated by $R_{ab}$\,. 
The diagonal generators $D, C, B$ are dilatation, central charge and hyper charge, respectively.  
The supertranslations $Q_{\al b}, \bar Q_{a \dot \be}$ and 
superconformal boosts $S_{a\be}, \bar S_{\dot \al b}$ are given by 
\begin{align}
&Q_{\al b}=E_{\al b}\,, \qquad \bar Q_{a \dot \be}=E_{a \dot \be}\,,
\qquad S_{a\be}=E_{a \be}\,,  \qquad  \bar S_{\dot \al b}=E_{\dot \al b}\,.  
\end{align}

\section{Constant classical $R$-matrix \label{app: classialR}}

We summarize here the notation of the classical $R$-matrix,
which is independent of the spectral parameter 
(For example, see \cite{CP}).

\subsection{Classical Yang-Baxter equation} 

Let $\alg{g}$ be a bosonic Lie algebra over $\mathbb{C}$. 
For $a_i, b_i \in \alg{g}$, an element denoted by 
\begin{align}
r=\sum_i a_i \otimes b_i \in \alg{g}\otimes\alg{g}
\end{align}
is called {\it classical $r$-matrix} if it satisfies the  
{\it classical Yang-Baxter equation (CYBE)}; 
\begin{align}
[r_{12},r_{13}]+[r_{13},r_{23}]+[r_{12},r_{23}]=0\,,  
\label{CYBE}
\end{align}
where the action of $r_{ij}$ is extended 
to three sites $\alg{g}\otimes\alg{g}\otimes\alg{g}$ such as   
\begin{align}
r_{12}=\sum_i a_i \otimes b_i  \otimes 1
\qquad 
r_{23}= \sum_i 1\otimes a_i \otimes b_i 
\qquad 
r_{13}=\sum_i a_i \otimes 1\otimes b_i\,.   
\end{align}

\medskip

Suppose that there exists  the invariant non-degenerate symmetric
bilinear form $\lan ~, ~\ran $ on $\alg{g}$\,. 
With the bilinear form, the linear operator $R:\alg{g}\to \alg{g}$ can be introduced though
the following relation; 
\begin{align}
R(X)=\lan r , 1\otimes X \ran =\sum_i a_i \lan b_i, X \ran \in \alg{g}  
\label{RR12}
\end{align}
for any $X\in \alg{g}$\,. This operator $R$ is also referred as to the classical $R$-matrix. 
With this notation, CYBE \eqref{CYBE} is equivalent to 
\begin{align}
[R(X),R(Y)]-R([R(X),Y]+[X,R(Y)])=0\,, 
\label{RCYBE}
\end{align}
if and only if the $r$-matrix is skew-symmetric;
\begin{align}
r_{21}=\sum_i b_i\otimes a_i =-r\,. 
\end{align}
Indeed, noting the following relations for any $X, Y\in \alg{g}$\,, 
\begin{align}
[R(X),R(Y)]&= \lan [r_{12},r_{13}] ,1\otimes X\otimes Y\ran\,, \no \\
-R([R(X),Y])&= \lan [r_{13},-r_{32}] ,1\otimes X\otimes Y\ran\,, \no \\
-R([X,R(Y)]&= \lan [r_{12},r_{23}] ,1\otimes X\otimes Y\ran \,, 
\end{align}
one can see that the relation \eqref{RCYBE} is nothing but \eqref{CYBE} if $R$ is skew-symmetric. 

\medskip

Here it is worth mentioning the generalization of CYBE \eqref{RCYBE} such as  
\begin{align}
[R(X),R(Y)]-R([R(X),Y]+[X,R(Y)])=-c^2[X,Y]\,, 
\label{mRCYBE}
\end{align}
for any $X,Y\in \alg{g}$ with a complex parameter $c\in \mathbb{C}$\,. 
The relation (\ref{mRCYBE}) is called {\it the modified classical Yang-Baxter equation} (mCYBE).   
The standard examples of the classical $r$-matrix (or $R$-matrix) satisfy CYBE \eqref{CYBE} 
(or \eqref{RCYBE})\,,  
while (twice of) the skew-symmetric parts of them satisfy mCYBE \eqref{mRCYBE}.

\subsection{Skew-symmetric $r$-matrix for $\alg{gl}(M|N)$ \label{app:glmn}}

Let us summarize typical constant $r$-matrices for the 
Lie superalgebra $\alg{gl}(M|N)$\,. 
The Lie superalgebra $\alg{gl}(M|N)$ is $(M+N)^2$-dimensional algebra over $\mathbb{C}$ 
and generated by $E_{ij}$ with $i,j = 1,\cdots ,M+N$ satisfying the relations; 
\begin{align}
[E_{ij},E_{kl}] = \de_{kj} E_{il} - \de_{il} E_{kj} (-1)^{(\bar i+\bar j)(\bar k+\bar l)}\,. 
\label{glnm}
\end{align}
Here the parity of indices are defined as 
$\bar i=0$ for $i=1, \cdots , M$ and 
$\bar i=1$ for $i=M+1, \cdots , N+M$. 

\medskip 

There are three typical solutions of (m)CYBE. 
The first one is the trivial solution $r=0$ for CYBE. 
The second one is the classical $r$-matrix $r_{\rm DJ}$ of Drinfeld-Jimbo type 
\cite{Drinfeld1,Drinfeld2,Jimbo}
\begin{eqnarray}
r_{\rm DJ}=c \sum_{1\leq i<j \leq M+N} E_{ij}\wedge E_{ji} (-1)^{{\bar i}\,{\bar j}}\,. 
\end{eqnarray} 
This is a solution of mCYBE.  

\medskip 

The third solution is the non-standard classical $r$-matrix $r_{\rm tw}^{(i,i)}$, 
which are obtained by twisting $r_{\rm DJ}$ with a root generator $E_{ij}$\,. 

\medskip 

The twists by the bosonic roots $E_{\al\be}$ and $E_{ab}$ with $\alpha < \be$ and $a<b$ ($\al, \be =1,\cdots, M$ and $a, b=M+1, \cdots, N+M$)
are given by 
\begin{align}
&r_{\rm tw}^{(\alpha,\beta)} \equiv [\Delta(E_{\al \be}), r_{\rm DJ}]
= c \left[(-E_{\al\al}+E_{\be\be})\wedge E_{\al \be}
-2 \sum_{\kappa=\al+1}^{\be-1} E_{\al\kappa}\wedge E_{\kappa \be}\right]\,, \nonumber  \\
& r_{\rm tw}^{(a,b)} \equiv [\Delta(E_{ab}), r_{\rm DJ}]
= c\left[(E_{aa}-E_{bb})\wedge E_{ab} + 2 \sum_{k=a+1}^{b-1} E_{ak}\wedge E_{kb}\right]\,,  \nonumber 
\end{align}
where the coproduct is defined in \eqref{copro}. 
These are solutions of CYBE rather than mCYBE. 
We will call them the bosonic twists. 

\medskip 

The fermionic root $E_{\al,b}$ gives rise to 
\begin{eqnarray}
r_{\rm tw}^{(\alpha,b)} \equiv [\Delta(E_{\al b}), r_{\rm DJ}]
=c \left[(E_{\al\al}+E_{bb})\wedge E_{\al b}
+2 \sum_{\kappa=1}^{\al-1} E_{\al\kappa}\wedge E_{\kappa b} 
-2\sum_{k=a+1}^{M+N } E_{\al k}\wedge E_{k b}\right]\,, \nonumber 
\end{eqnarray}
where $c$ is a Grassmann odd parameter rather than a complex number, 
so that the $r$-matrix should be Grassmann even \cite{Tolstoy}. 
We will refer the twists as to the fermionic twists. 
However it does not seem to be a solution of (m)CYBE. 

\medskip 

An interesting example of the fermionic twist is given by $E_{1,M+N}$ as follows: 
\begin{align}
&r_{\rm tw}^{(1,M+N)}=[\Delta(E_{1,M+N}), r_{\rm DJ}]=-c E_{1,M+N} \wedge (E_{11}+E_{M+N,M+N})\,. 
\end{align}
This is a solution of CYBE. 
When $M=N=4$, it reduces to \eqref{fer18}\,.

\end{document}